\documentclass[journal=jacsat,manuscript=article]{achemso}

\usepackage{graphicx}% Include figure files
\usepackage{dcolumn}% Align table columns on decimal point
\usepackage{bm}% bold math
\usepackage[mathlines]{lineno}% Enable numbering of text and display math
\usepackage{etoolbox}
\usepackage[usenames,dvipsnames]{color}
\usepackage{gensymb}
\usepackage{graphicx}
\usepackage{subfigure}
\usepackage{chemformula}
\usepackage{amsmath}
\usepackage{amssymb}
\usepackage{dsfont}
\usepackage{float}
\usepackage{hyperref}
\usepackage{multirow}
\usepackage{rotating}
\usepackage{epstopdf}
\usepackage{textcomp}
\usepackage{gensymb}
\usepackage{color}
\usepackage{wrapfig}
\usepackage{fancyhdr}
\usepackage{lipsum}
\usepackage{tabularx}
\usepackage[normalem]{ulem}
\usepackage{soul}
\usepackage[capitalize]{cleveref}

\usepackage[T1]{fontenc}
\DeclareUnicodeCharacter{2212}{-}
\DeclareUnicodeCharacter{03B4}{$\delta$}

\usepackage[version=3]{mhchem} % Formula subscripts using \ce{}
\usepackage{chemformula}
\author{Sk Kalimuddin}
\affiliation{School of Physical Sciences, Indian Association for the Cultivation of Science, 2A \& 2B Raja S C Mullick Road, Jadavpur, Kolkata 700 032, India}

\author{Biswajit Das}
\affiliation{Department of Basic Science and Humanities, Dr. Sudhir Chandra Sur Institute of Technology \& Sports Complex, 540, Dum Dum, Surer math, Kolkata 700074, India}
\altaffiliation{School of Physical Sciences, Indian Association for the Cultivation of Science, 2A \& 2B Raja S C Mullick Road, Jadavpur, Kolkata 700 032, India}

\author{Sudipta Chatterjee}
\affiliation{Department of Condensed Matter and Materials Physics, S. N. Bose National Centre for Basic Sciences, Kolkata 700106, India}

\author{Arnab Bera}
\affiliation{School of Physical Sciences, Indian Association for the Cultivation of Science, 2A \& 2B Raja S C Mullick Road, Jadavpur, Kolkata 700 032, India}

\author{Satyabrata Bera}
\affiliation{School of Physical Sciences, Indian Association for the Cultivation of Science, 2A \& 2B Raja S C Mullick Road, Jadavpur, Kolkata 700 032, India}

\author{Kalyan Kumar Chattopadhyay}
\affiliation{Thin Film and Nanoscience Lab, Department of Physics, Jadavpur University, Kolkata 700032, India}
\email{kalyan_chattopadhyay@yahoo.com}

\author{Mintu Mondal}
\affiliation{School of Physical Sciences, Indian Association for the Cultivation of Science, 2A \& 2B Raja S C Mullick Road, Jadavpur, Kolkata 700 032, India}
\email{Mintu.Mondal@iacs.res.in}

\title[An \textsf{achemso} demo]{Barrier height inhomogeneity and origin of 1/f-noise in topological insulator-based photo-detector}

\abbreviations{IR,NMR,UV}

\begin{document}

%\begin{tocentry}
%Some journals require a graphical entry for the Table of Contents.
%This should be laid out ``print ready'' so that the sizing of thetext is correct.
%Inside the \texttt{tocentry} environment, the font used is Helvetica
%8\,pt, as required by \emph{Journal of the American Chemical Society}.

%The surrounding frame is 9\,cm by 3.5\,cm, which is the maximum permitted for  \emph{Journal of the American Chemical Society}
%graphical table of content entries. The box will not resize if the content is too big: instead it will overflow the edge of the box.

%This box and the associated title will always be printed on a
%separate page at the end of the document.

%\end{tocentry}


\begin{abstract}
Topological insulators (TIs) with symmetry-protected surface states, offer exciting opportunities for next-generation photonic and optoelectronic device applications. The heterojunctions of TIs and semiconductors (e.g. Si, Ge) have been observed to excellent photo-responsive characteristics. However, the realization of high-frequency operations in these heterojunctions can be hindered by unwanted 1/f (or Flicker) noise and phase noise. Therefore, an in-depth understanding of 1/f noise figures becomes paramount for the effective utilization of such materials.Here we report optoelectronic response and 1/f noise characteristics of a p-n diode fabricated using topological insulator, \ch{Bi2Se3} and silicon for potential photo-detector. Through meticulous temperature-dependent current-voltage (I-V) and capacitance-voltage (C-V) measurements, we ascertain crucial parameters like barrier height, ideality factor, and reverse saturation current of the photodetector. The low-frequency 1/f conductance noise spectra suggest a significant presence of trap states influencing the optoelectronic transport properties. The forward noise characteristics exhibit typical 1/f features, having a uni-slope across four decades of frequency, suggesting a homogeneous distribution of barrier height. The spectral and photocurrent-dependent responses show the power law behavior of noise level on photon flux.  The hybrid heterojunction demonstrates excellent photo-response and reasonably low noise level, promising signatures for the room-temperature visible photodetector applications.
\end{abstract}
\renewcommand\thefootnote{}
\footnotetext{\textbf{Abbreviations:} BH, barrier height; LFCNS, Low frequency conductance noise spectroscopy; PSD, power spectral density.}

\section{Introduction}\label{sec1}
Topological insulators (TIs), a family of nontrivial quantum matter with insulating bulk states and time-reversal symmetry-protected Dirac-like surface states, have been widely used in thermoelectrics, photonics, and cutting-edge optoelectronic devices \cite{zhang2009topological,zhang2015improved,hong2016n,mu2015graphene,hajlaoui2014tuning,zheng2015patterning}. Among these various TIs, Bismuth selenide (\ch{Bi2Se3}) has been identified as a phenomenal TI candidate primarily because of its unique features, such as a small bandgap of approximately 0.35 eV\cite{Wang2018_Bi2Se3flakes_photodetector}, high surface mobility ($\sim$ 10$^4$ cm$^{2}$V$^{-1}$s$^{-1}$), and outstanding photo-conductivity \cite{sharma2016high,yang2018tuning,yan2013large}. Moreover, \ch{Bi2Se3} is stable under ambient conditions, and the robust characteristics of Dirac fermions can be observed even at room temperature \cite{zhang2016high}. Nevertheless, the practical uses of \ch{Bi2Se3} have been severely limited by the relatively reduced responsivity and the weak helicity-dependent photocurrent found in \ch{Bi2Se3}-based photoelectronic devices \cite{mciver2012control,yan2014topological}.

On the other hand, incorporating TIs and semiconductors (like Si, Ge, or GaN) to develop heterojunctions has been seen as an effective way to overcome the limitations of \ch{Bi2Se3} \cite{zhang2016high,das2017topological,lu2023bi2te3,zeng2023self,yin2022self}. Heterojunctions of TIs and semiconductors manifest excellent optoelectronic properties and have been frequently employed in the past few years to prepare high-performance optoelectronic devices\cite{zhang2016high,das2017topological,lu2023bi2te3,zeng2023self}. For instance, the \ch{Bi2Se3}/Si (100) heterojunction, prepared via thermal evaporation, demonstrates high responsivity and detectivity \cite{zhang2016high}. The \ch{Bi2Se3} nanowire/Si heterostructure, synthesized via the vapor-liquid-solid method, also shows high photoresponsive properties\cite{Liu.VLS}. In another work, \ch{Bi2Se3} nano-flake/Si-nanowire-based heterojunction also displays excellent photo responsivity and detectivity \cite{das2017topological}. Despite numerous optoelectronic studies in \ch{Bi2Se3}/Si heterojunctions, a low-frequency 1/f noise study along with the optoelectronic properties to understand the charge transport mechanism at the interface in this exotic heterojunction has not been conducted so far.

The low-frequency 1/f noise (or flicker noise), a significant performance factor in electronic devices for practical applications, has been found to be a useful technique for studying charge transport mechanisms \cite{weissman19881,Raychaudhuri2002_basic,chatterjee2021emergence,bisht2022disorder,kalimuddin2023macroscopic,SatyaPolymer,ArnabMoTe2}. It has now been well recognized that the source of 1/f noise can differ across materials as a result of fluctuations in number density or mobility orders \cite{Dutta1981_noise_review}. In light of technological applications, the magnitude of 1/f noise is of extreme importance for any electronic device since it can mix with phase noise and restrict the device's performance during high-frequency communications \cite{karnatak20171,islam2023benchmarking}. Therefore, in order to use a new material system for practical applications, it is imperative that the origin and amplitude of 1/f noise be thoroughly understood. Consequently, the 1/f noise properties of \ch{Bi2Se3}/Si interfaces play an important role in the incorporation of \ch{Bi2Se3}/Si into technological applications. Low-frequency noise spectroscopy studies in \ch{Bi2Se3} (both in bulk and thin films) have been reported before\cite{Bi2Se3.noise.acsnano,Bi2Se3.noise.pss,Bi2Se3.noise.film}, however, understanding the charge transport mechanism in this novel \ch{Bi2Se3}/Si heterostructure via 1/f noise spectroscopy is yet to be investigated.

In this work, we report the optoelectronic properties and low-frequency 1/f noise of a chemical bath-deposited \ch{Bi2Se3}/Si diode down to 200 K. The photodetector device exhibits good photoresponse which makes it a promising candidate for photodetection in the visible spectrum range. Detailed temperature dependent noise measurement reveals that the 1/f noise floor arises from large barrier inhomogeneity. From the intensity-dependent optical response study, we establish that \ch{Bi2Se3}/Si has fairly high responsivity and detectivity with a fast response speed making it a promising candidate for room temperature photodetector application.

\section{Results \& Discussion}\label{sec2}
\begin{figure*}[t]
\centering
\includegraphics[width=1.0\textwidth]{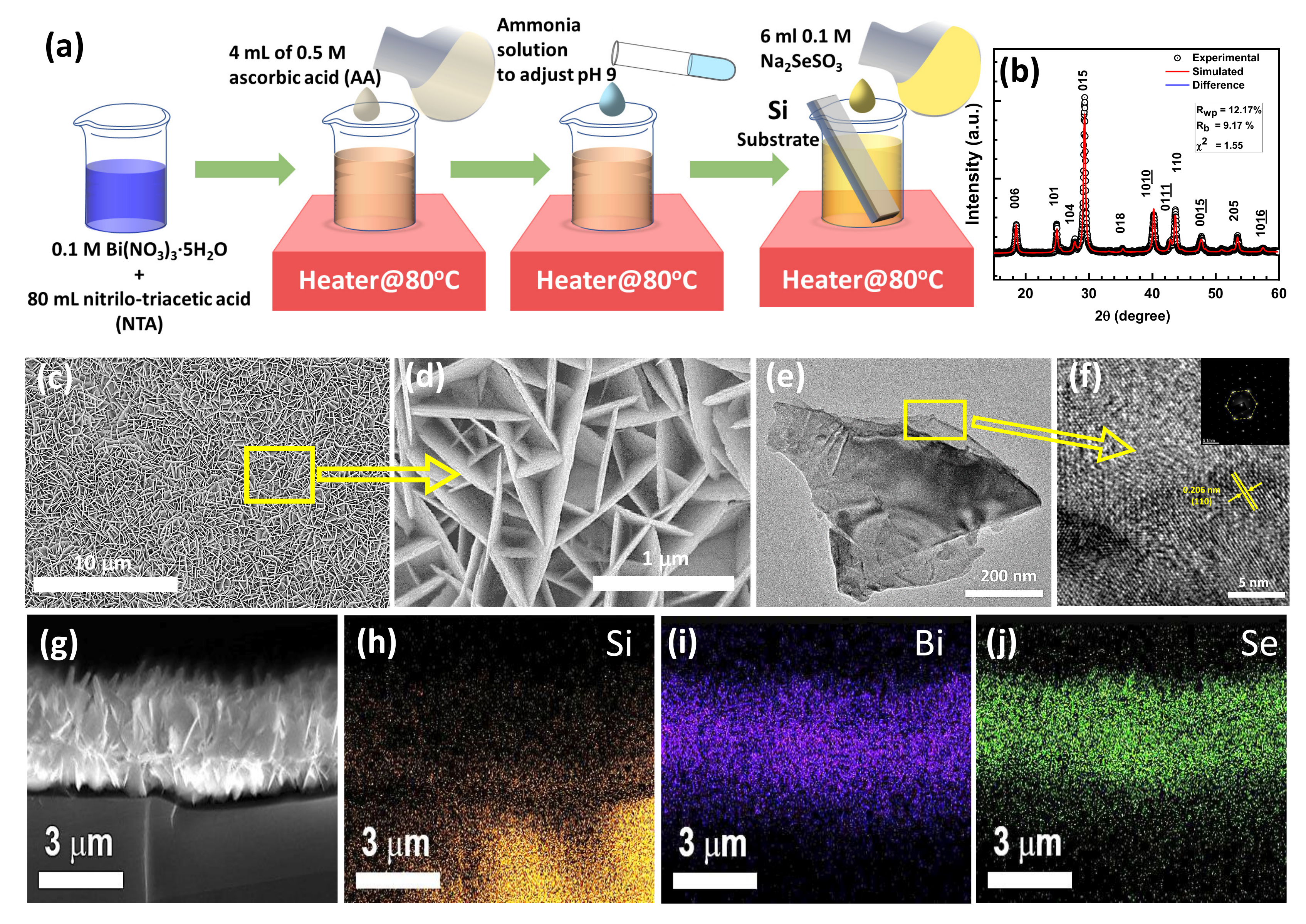}
\caption{\textbf{Synthesis and Morphological Characterization} (a) Synthesis schematic of n-type \ch{Bi2Se3} film deposition on p-type Si(111) substrate. (b) X-ray diffraction pattern of \ch{Bi2Se3} thin film with Rietveld refinement. (c) Low and (d) high magnification SEM micrograph of \ch{Bi2Se3} nano-flakes. (e) TEM micrograph of a single \ch{Bi2Se3} nano-flake and its (f) HRTEM image nano-flake. Inset is the corresponding hexagonal SAED pattern. (g) The cross-sectional SEM image of the p-n junction where (h-j) represents the elemental mapping for Si, Bi, and Se, respectively.}
\label{Fig:Characterization}
\end{figure*}

The growth of the precursors and device fabrication process has been schematically highlighted in Fig [\ref{Fig:Characterization}(a)]. To investigate the crystallinity and single phase formation, \ch{Bi2Se3} thin film which is deposited on Si substrate has been characterized using X-ray diffraction in the 2$\theta$ range from 15 - 60$\degree$. Characteristic XRD peaks of \ch{Bi2Se3} film at room temperature are shown in Fig. [\ref{Fig:Characterization}b]. Rietveld refinement of the corresponding XRD pattern confirms the single-phase formation of the sample. \ch{Bi2Se3} crystalizes in the trigonal crystal structure with (R-3m:H) space group. All the indexed peaks correspond to the trigonal structure of \ch{Bi2Se3} with lattice parameters a= 4.1482$~$\AA$~$and c = 28.673$~$\AA$~$, which matches with peaks reported earlier (JCPDS 33-0214)\cite{JCPDS.crystal.1963}. 

The morphological studies and composition of the deposited film have been obtained via SEM and a low magnification surface image with a scan area of 25 x 20 $\mu m^2$ is shown in Fig. [\ref{Fig:Characterization}c], and it is conspicuous that surface of the film is quite uniform and \ch{Bi2Se3} flakes are uniformly distributed over a large area. A high-resolution surface image of the same is shown in Fig. [\ref{Fig:Characterization}d]) where it is clear that \ch{Bi2Se3} flakes are uniformly distributed. % like the petals of a flower. The flakes have a larger surface area compared to bulk crystals. %High-resolution transmission electron microscopy reveals micro-structural information even up to the atomic scale including the reciprocal space. 
Flakes resemble petals-like structures with flattened edges. \ch{Bi2Se3} flakes are of the order of a few hundreds of nanometers. Fig [\ref{Fig:Characterization}e] shows a single flake of dimension 200 nm. %Although \ch{Bi2Se3} samples are good electron beam transparent, for good quality TEM imaging area was selected so that TEM beam is supposed to pass through flake edges with relatively lower thickness. 
The selected area electron diffraction (SAED) pattern has been captured from the flake microstructure and is shown in the inset of Fig. [\ref{Fig:Characterization}f]. From the SAED pattern, it is crystal clear that \ch{Bi2Se3} flakes in reciprocal space exhibit a crystalline structure.% having a \textcolor{red}{hexagonal symmetry of a rhombohedral phase(space group: R3m).}

\subsection{Current-voltage characteristics}
\begin{figure*}[t]
\centering
\includegraphics[width=1.0\textwidth]{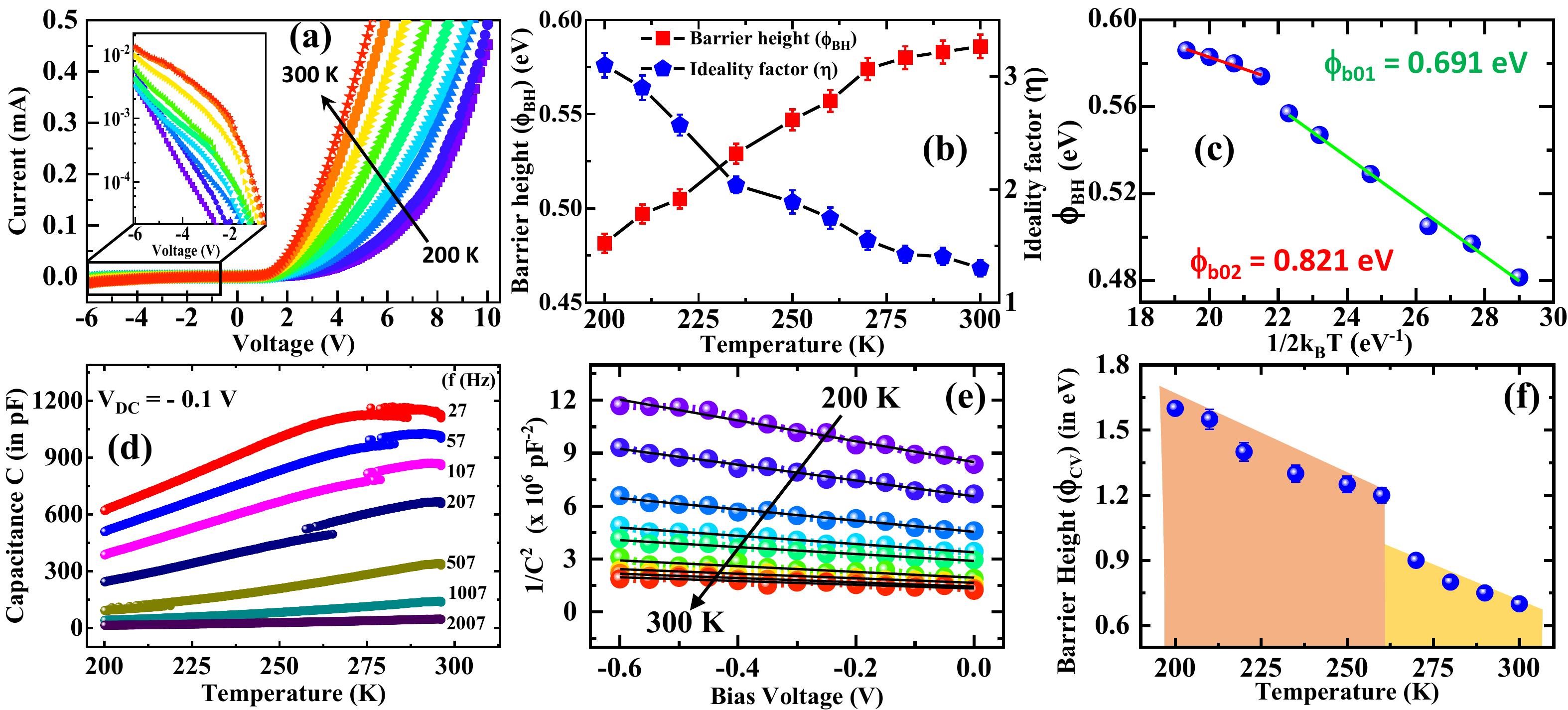}
\caption{\textbf{Electrical characterizations: I-V spectra and junction capacitance.} (a) Temperature-dependent I-V characteristics of Si- \ch{Bi2Se3} p$-$n junction. The inset presents semi-log I$-$V characteristics in reverse bias. (b) Temperature-dependent barrier height (left) and ideality factor(right), determined from IV characteristics. (c) Barrier height vs $\frac{1}{2~k_BT}$ plot shows two-slope behaviour with their corresponding mean of $\phi_{b0}$. (d) Temperature and frequency-dependent junction capacitance (C). (e) Variation in $(1/C^2)$  and (f) Barrier height ($\phi_{b0}$) obtained from CV measurements.}
\label{Fig:IVCV}
\end{figure*}

To understand the nature of electrical transport at \ch{Bi2Se3}/Si p-n interface current-voltage (I-V) characteristics has been performed in the temperature range of 150–300 K. Fig. [\ref{Fig:IVCV}a] shows the I-V characteristics of \ch{Bi2Se3}/Si. The diode is fairly rectifying under reverse bias with an ideality factor of (1.3) at 300 K. The junction parameters are extracted using the thermionic theory given by Eq. [\ref{Eq:diode_current}],
\begin{equation}
I = I_s\Bigl(e^{qV/\eta KT}-1\Bigl)
\label{Eq:diode_current}
\end{equation}
Where, I$_s$ = AA$^*T^2e^\frac{\phi_{b0}}{KT}$ \cite{Chen2011_Rev_sat_current}, is the reverse saturation current, $\eta$ is the ideality factor, V is the applied bias, T is the absolute temperature, q is the electronic charge, and K is the Boltzmann constant. In reverse saturation current (I$_s$) A is the effective contact area, $A^*$ is the effective Richardson coefficient (32 Acm$^{-2}$K$^{-2}$)\cite{Hacke1993_Richardson_pSi_2,Cankaya2004_Richardson_3} for p-Si and $\phi_{b0}$ is the apparent zero bias barrier height. Ideality factor ($\eta$) is determined from the slope of the linear fitted I-V plot using the relation, $\eta = \frac{q}{2.303kT}\frac{dV}{d(logI)}$, and barrier height ($\phi_{bo}$) is estimated using the relation, $\phi_{bo}= \frac{kT}{q}ln(\frac{AA^*T^2}{I_s})$.

The temperature-dependent plot of the barrier height $(\phi_{bo})$ (see Fig [\ref{Fig:IVCV}(b,c)]) shows a decrease in barrier height with the decrease in temperature, this variation in barrier height can be attributed to Silicon/\ch{Bi2Se3} interface inhomogeneity and the increase in ideality factor with lowering the temperature is anticipated to the non-epitaxial deposition of \ch{Bi2Se3} film on Si. The non-epitaxial deposition of the film results in an atomically flat but rough interface which further results in local variation of the electric field due to external bias and causes barrier height to vary locally in the interface region. %This distribution of barrier height leads to flow current dominantly through patches of lower barrier height. 
Usually, $\eta \sim 2$ is attributed to strong recombination away from the interface. Furthermore, the values of $\eta$ $\geq$ 2, especially in low temperatures, suggest dominant recombination generation near the interface.

%At room temperature, conduction electrons have sufficient thermal energy and barrier height is lower hence conduction of electrons becomes feasible. With the decrease in temperature the barrier height of the junction increases and also has a distribution locally at the interface in nano-meter scale. Thereby low temperature transport of carriers is dominated through patches of low barrier height channels.

The plot of $\phi_{b0}$ vs 1/$2~k_BT$ shows two-slope behaviour (see [\ref{Fig:IVCV}c]), where $\phi_{b01}$ \& $\phi_{b02}$ are the mean values of the Gaussian local distribution of barrier heights. In 1991, J. H. Werner and H. H. G\"uttler, proposed an analytical potential fluctuation model\cite{Werner1991_Barrierihmogogeneity} to describe the charge transport across a Schottky barrier contact and observation of idealities $\eta > $1 as well as its temperature dependence. Here the spatial barrier heights follow a Gaussian distribution given by Eq. [\ref{Eq:BH1}],\cite{Werner1991_Barrierihmogogeneity}

\begin{equation}
P(\phi_{b0})=\frac{1}{\sqrt{2\pi\sigma_s^2}}e^{-\frac{(\overline{\phi_{b0}}-\phi_{b0})^2}{\overline{2\sigma_s^2}}}
\label{Eq:BH1}
\end{equation}
Where $\overline{\phi_{b0}}$,$\phi_{b0}$ are zero bias mean and local barrier height respectively. $\sigma_s$ is the standard deviation of the distribution It gives a quantitative measure of the extent of barrier inhomogeneity. The total current across the junction is integrated for all P($\phi_{b0}$) which results in temperature dependence of the barrier height as $\phi_{b0}= \overline{\phi_{b0}} -\frac{\sigma_s^2}{2kT}$. 

Therefore, quantification of $\sigma_s$ becomes trivial from the $\phi_{b0}$ vs $\frac{1}{2kT}$ plot. %Equation [\ref{Eq:BH1}] suggests that $\sigma_s$ of barrier height can be determined from the $\phi_{b0}$ vs $\frac{1}{2kT}$ plot. 
It is interesting to note that, the plot of Fig. [\ref{Fig:IVCV}c] shows two distinct slopes in two different regimes, indicating the presence of two Gaussian distributions. Such multi Gaussian feature was well addressed direct measurement of interface barrier potential \cite{vanalme1997ballistic,vanalme1999ballistic}. For the present device in the temperature region (270-300 K), ($\sigma_s$) and ($\overline{\phi_{b0}}$) are found to be 73 meV and 0.693 eV respectively whereas in the low-temperature region (210-260 K) ($\sigma_s$) and ($\overline{\phi_{b0}}$) is found to be 108 meV and 0.819 eV respectively. Therefore, $\sigma_s$ for temperatures 270-300 K suggests barrier height is small and hence diode behaves close to the ideal nature with smaller values of ideality factor.

\subsection{Capacitance-voltage characteristics}
To understand the interplay of barrier inhomogeneity and charge separation at the junction, frequency-dependent Capacitance has been measured down to 200 K for different small DC biases.  Temperature-dependent capacitance is shown in Fig. [\ref{Fig:IVCV}a] for frequencies 57-20007 Hz. The bias dependent depletion-layer transition capacitance $C_j$ can be given by the equation\cite{Maeda2017_C_bias1, Reddy2020_Cbias_3, Tripathi2012_C_bias2},
\begin{equation}
	\frac{1}{C_j^2} = \frac{2(V_{bi}+V)}{q\epsilon_s\epsilon A^2(N_d-N_a)} = \frac{\epsilon_s}{\omega}
 \label{Eq:CV}
\end{equation}
where $V_{bi}$ is built-in potential, V is the applied bias, $\epsilon $ is the dielectric constant of \ch{Bi2Se3} (15.06) \cite{Tse2015_Dielectric_cinstant}, A is the area of the junction of contact or electrode ($N_d- N_a$) is the net donor carrier concentration. 

The temperature-dependent capacitance plot shows two levels of capacitance value over a smaller window of temperatures earlier experimentally observed from ballistic electron emission miscroscopy\cite{vanalme1997ballistic,vanalme1999ballistic}. The temperature window of bistability has a substantial shift with frequency suggesting a cross-over between the two levels' barrier heights. This indicates the existence of two Gaussian distributions and two mean barrier height analytical potential fluctuations predicted\cite{Werner1991_Barrierihmogogeneity,vanalme1997ballistic}. 

\subsection{Benchmarking noise characteristics of the device}
\begin{figure*}[t]
\centering
\includegraphics[width=1.0\textwidth]{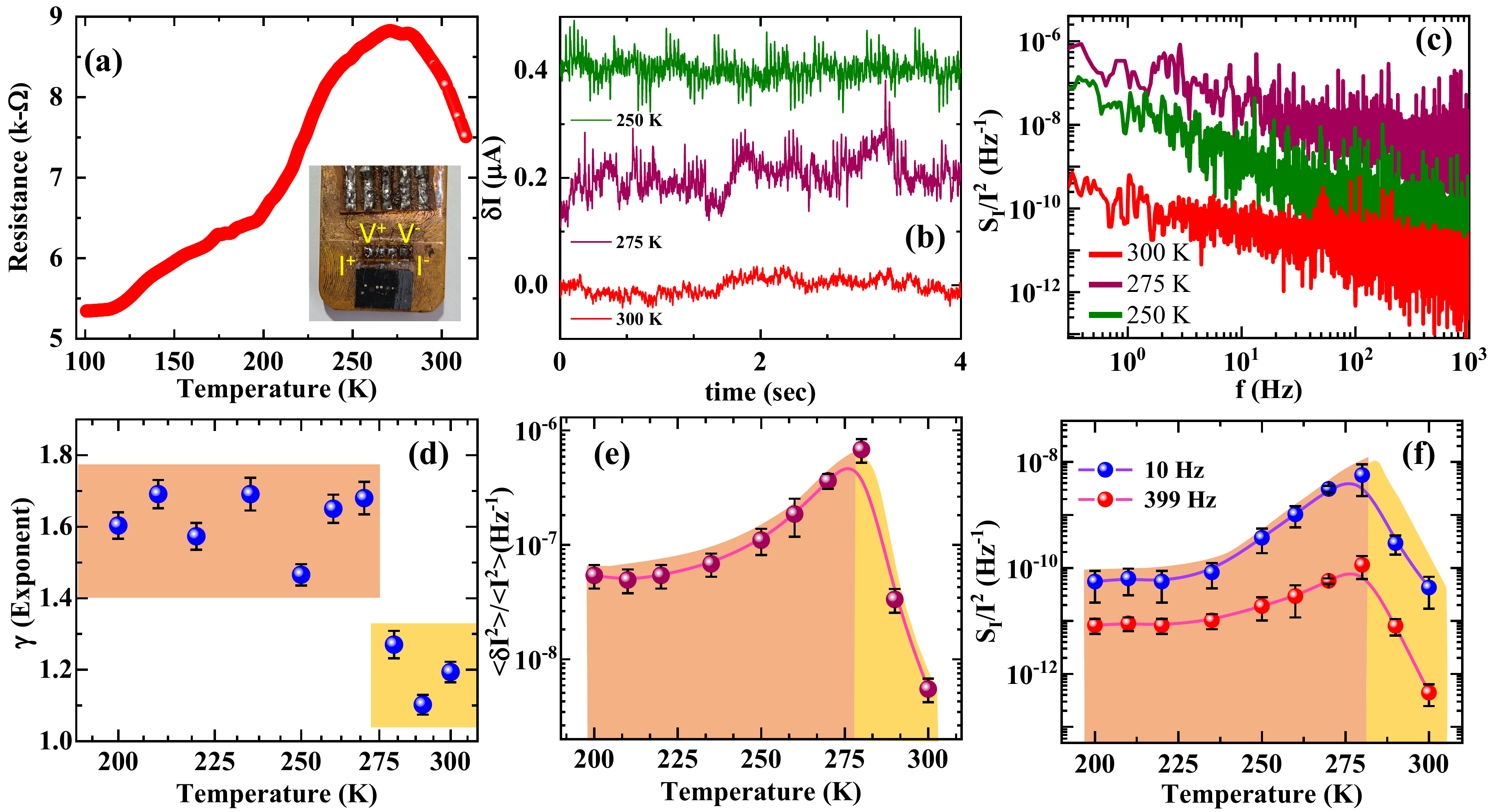}
\caption{\textbf{Low-frequency 1$/$f conductance noise.} (a) Resistance vs temperature of the \ch{Bi2Se3} film. (b) Representative time series of current fluctuations at fixed bias voltage. Vertically shifted by 0.2 $\mu$A for clarity. (c) Normalized PSD of corresponding current fluctuations.  (d) Temperature dependence of the noise exponent ($\gamma$), where $S_I/I^2 \propto 1/f^{\gamma}$.  (e) Relative variance vs temperature within the bandwidth [10$^{-1}$ - 10$^3$ Hz]. (f) Variation of the noise contributions to PSD at 10 Hz and 339 Hz.}
\label{Fig:Noise1}
\end{figure*}
Temperature-dependent four-point resistance is plotted in Fig. [\ref{Fig:Noise1}(a)]. A clear metallic nature of the film is observed suggesting significant conduction from surface states of \ch{Bi2Se3}. In addition, a broad hump is observed near 280~K. Here, Low-frequency conductance noise spectroscopy (LFCNS) is employed to understand the role of barrier inhomogeneity at the interface on device performance and its impact on external temperature. In the present study, a constant voltage is applied to the device with a series resistance ($R_S$), and the time series voltage is digitized at a sampling rate of $2^{10}$ Hz using a 24-bit analogue to digital converter card. The mean-subtracted voltage fluctuations ($\delta$v (t)= v(t)-$\langle v(t)\rangle$), which is further converted to current fluctuations using $\delta$I(t) = $\frac{\delta V(t)}{R_S}$. 

A plot of representative time series of current fluctuations is shown in Fig. [\ref{Fig:Noise1}(b)]. Each data has been vertically-shifted by 0.4 for clarity. The power spectral density (PSD) is computed from $\delta I(t)$ using Eq. [\ref{eq:PSD}],
\begin{equation}
S_I(f) = \lim_{T\to\infty} \frac{1}{T} \bigg| \int_{-T/2}^{+T/2}\delta I(t)e^{-i \omega t}dt\bigg|^2,
\label{eq:PSD}
\end{equation}

The normalized PSD at a few selected temperatures has been presented in Fig. [\ref{Fig:Noise1}(c)]. It is worth noting that the PSD exhibits typical 1/f behaviour (i.e. $S_I \propto \frac{1}{f^{\gamma}}$). The device has a maximum noise floor near 280~K. The exponent ($\gamma$) in Fig. [\ref{Fig:Noise1} (d)] is relatively higher in the lower temperatures compared to temperatures above 280~K. The relative variance $\bigl( \frac{\langle\delta{I^{2}} \rangle}{\langle I^{2}\rangle}= \int_{0.3}^{1000} \frac{S_{I}(f)}{I^{2}}df \bigr)$ has been presented in Fig. [\ref{Fig:Noise1}(e)] which again suggests the presence of large 1/f noise around 280~K where a broad hump has been observed. The noise contributions from the device at frequencies 10 Hz \& 399 Hz have been presented in Fig. [\ref{Fig:Noise1} (f)], which reconfirms the uniformity of the noise spectra over the entire spectrum.

The excess flicker noise might arise from the most commonly observed barrier inhomogeneity. Wherein, a macroscopically flat but atomically inhomogeneous interface results in local distribution in the electric field and charge distribution. Temperature-dependent I-V characteristics also revealed the inhomogeneous nature of the interface. Therefore, major contributions of current fluctuations comprise charge trapping, de-trapping at the interface surface states, and the inhomogeneous nature of \ch{Bi2Se3}/Si interface\cite{Guettler1990_Noise_herbert_werner, Werner1991_Barrierihmogogeneity}. Random charge trapping at the interface modulates the total charge at the interface, thus modifying $\phi_{b0}$ resulting in excess noise. Since the current is exponential with voltage, minimal fluctuations in the barrier height result in significant 1/f noise. The uniform nature of 1/f noise and absence of Lorentzian feature over at least four decades suggests a homogeneous distribution of barrier height. 

\subsection{Photodetection }
\begin{figure*}[t]
\centering
\includegraphics[width=0.75\textwidth]{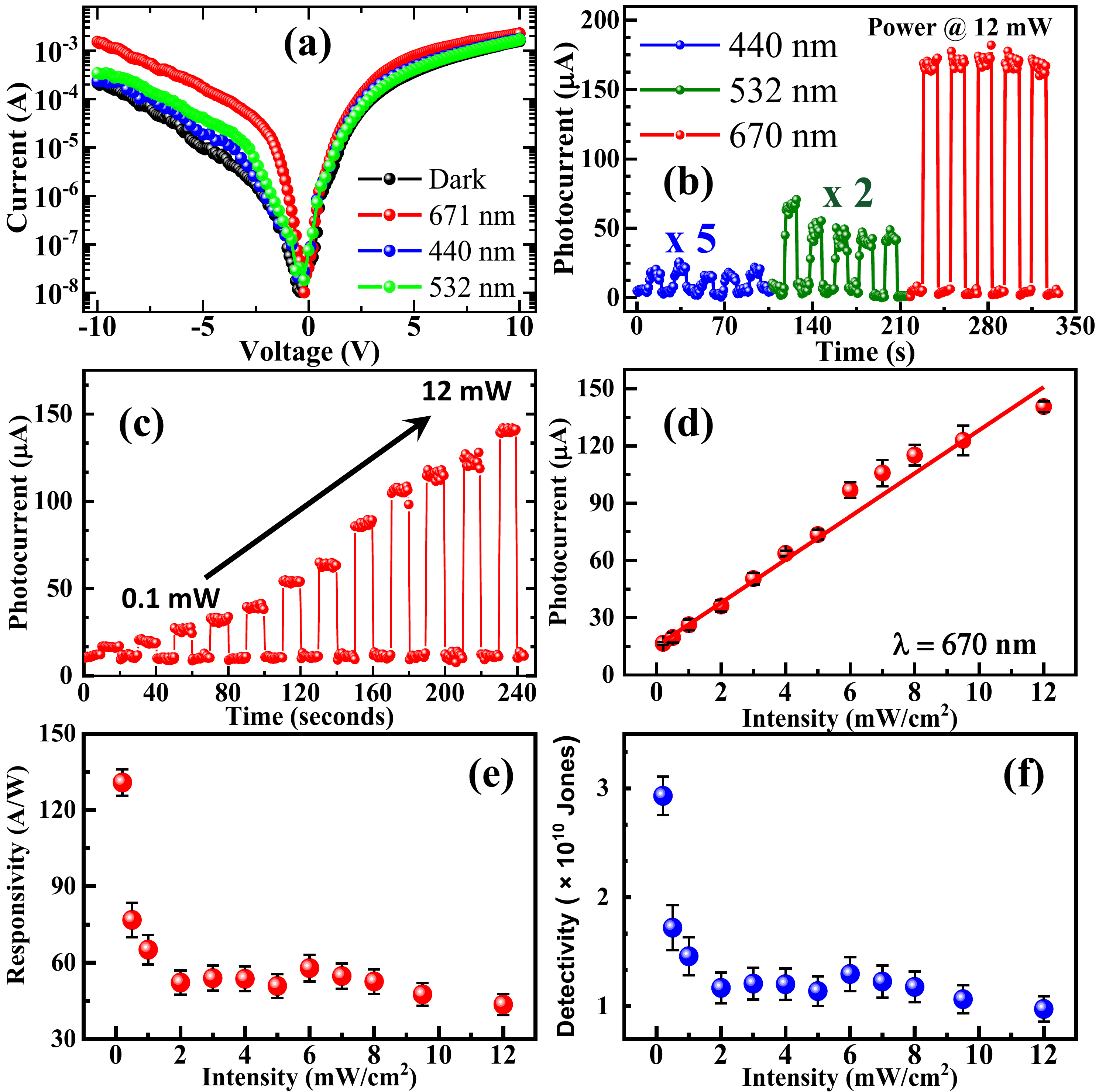}
\caption{\textbf{Photo response.} (a) Semi-log I-V characteristics under dark and light irradiated conditions at room temperature. (b) Optical switching of the device for different lasers at 12 mW laser power. (c)-(d) Variation of the photo-current (I$_{ph}$) with laser power (670 nm), where the solid line represents linear least square fit. Variation of (e) Responsibility and (f) detectivity using red laser.}
\label{Fig:Responsivity}
\end{figure*}

\ch{Bi2Se3}/Si p-n junction under reverse-biased shows remarkable rectification. To study the photo-response of the diode, the junction is irradiated with lasers of three different wavelengths, and IV measurements are carried out in the dark and with laser-irradiated conditions keeping intensities constant. Laser-dependent IV characteristics are shown in Fig [\ref{Fig:Responsivity}(a)]. Significant photocurrent generation can be observed for the Red laser (532 nm).
Photo-switching response under the same DC bias voltage for intensity of 12 $mW/cm^2$.with manual chopping is shown in Fig. [\ref{Fig:Responsivity}(b)]. The large photo-response of the heterojunction can be attributed to large photon harvesting at the interface states. Flakes of \ch{Bi2Se3} with large surfaces accelerate the photocurrent. 

Keeping in mind the maximum photon-harvesting near 670 nm, the detailed intensity-dependent photo-response has been studied for red laser only. Photo-current as a function of laser power (P) is shown in Fig. [\ref{Fig:Responsivity}(c) \& (d)]. Power-law variation of photocurrent ($I_{ph} \propto P^{\alpha}$), where $\alpha \sim$ 0.82 consistent with previous reports\cite{Jia2020_intensity_powerlaw}.

The responsivity (R) of the device has been calculated using $R = \frac{I_p-I_d}{P_{in}}$\cite{Zhang2017_Photoresponse_GaAs} and is shown in Fig. [\ref{Fig:Responsivity}(e)] and a maximum responsivity of R=130 A/W is achieved at 0.1 $mW/cm^2$. Similarly, the detectivity (D) has been calculated using $D = R \frac{\sqrt{A}}{\sqrt{2qI_d}}$\cite{das2017topological}. A maximum of 3 x $10^{10}$ Jones is observed for 0.1 $mW/cm^2$.

\subsubsection{Photo switching}
\begin{figure*}[h]
\centering	\includegraphics[width=0.7\textwidth]{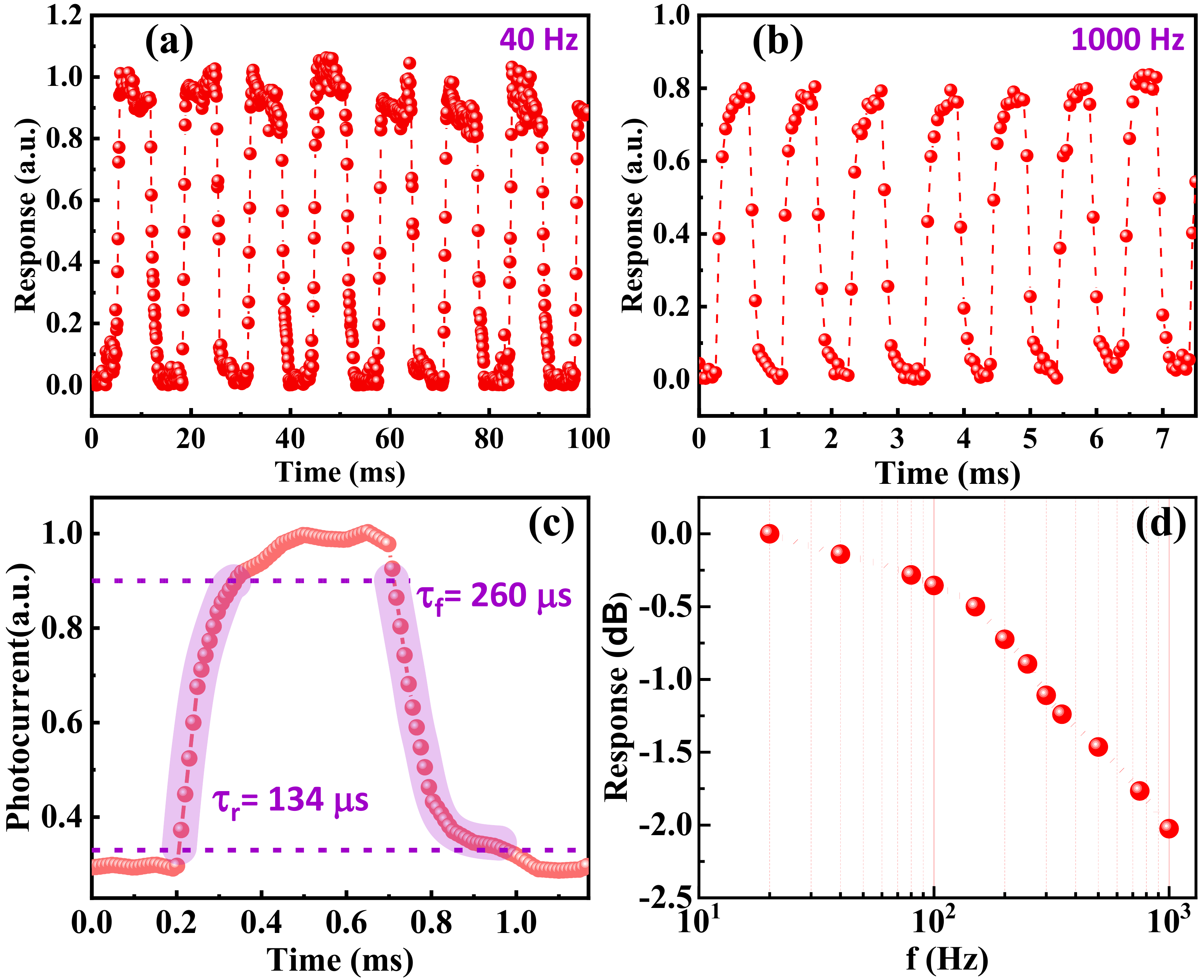}
\caption{\textbf{Photo-switching and response time.} (a)-(b) Photo-switching response at chopper frequencies of 40 and 1000 Hz respectively. (d) An expanded single cycle of photo-switching with corresponding rise and fall time. (e) Normalized photo response with frequency in dB scale having 2dB cut-off frequency of 900 Hz.}
\label{Fig:Photoswitch}
\end{figure*}
The quantitative estimates of the photo-detection indicate its possible use as a visible photo-detector. However, the response speed of the detector determines the ability of the detector to follow the fast varying optical signals, the capacity of information processing, and the imaging frame rate. This response speed is an important parameter for application in telecommunication purposes. For investigating the temporal response, the device is irradiated with an intermittent manual chopping. Fig [\ref{Fig:Photoswitch}(a)-(c)] shows the fast, stable, and reproducible photo-response of the device at chopping rates of 40, 500, and 1000 Hz respectively. The device responds with excellent stability and reproducibility over a wide range of frequencies for numerous cycles which indicates its ability to detect fast-varying optical signals and its potential use as a photodetector.  

The response time scales are calculated by analyzing the rise ($\tau_r$, the time interval from 10\% to 90\% of the maximum photocurrent) and fall time ($\tau_f$, the time interval from 90\% to 10\% of the maximum photocurrent) at rising and falling edges of a single cycle of the detected chopped signal. A single magnified and normalized photo-switching cycle at maximum chopping ($10^3$ Hz) is shown in Fig [\ref{Fig:Photoswitch}(c)] having a rise/fall time of  134/264 $\mu$s.

Nearly double the value of $\tau_f$ compared to $\tau_r$ is observed which can be understood trap-assisted mechanism. In the rising edge, photo-carries fill the trap states, and on saturation of filling the photo-current reaches its maximum value. Whereas on the falling edge, the carriers associated with trap states are released slowly giving rise to larger $\tau_f$. Moreover, the device exhibits a highly reproducible response even after consecutive thousand kilocycles of operation.

\subsection{Photon flux and excess noise}
\begin{figure}[t]
\centering
\includegraphics[width=0.85\columnwidth]{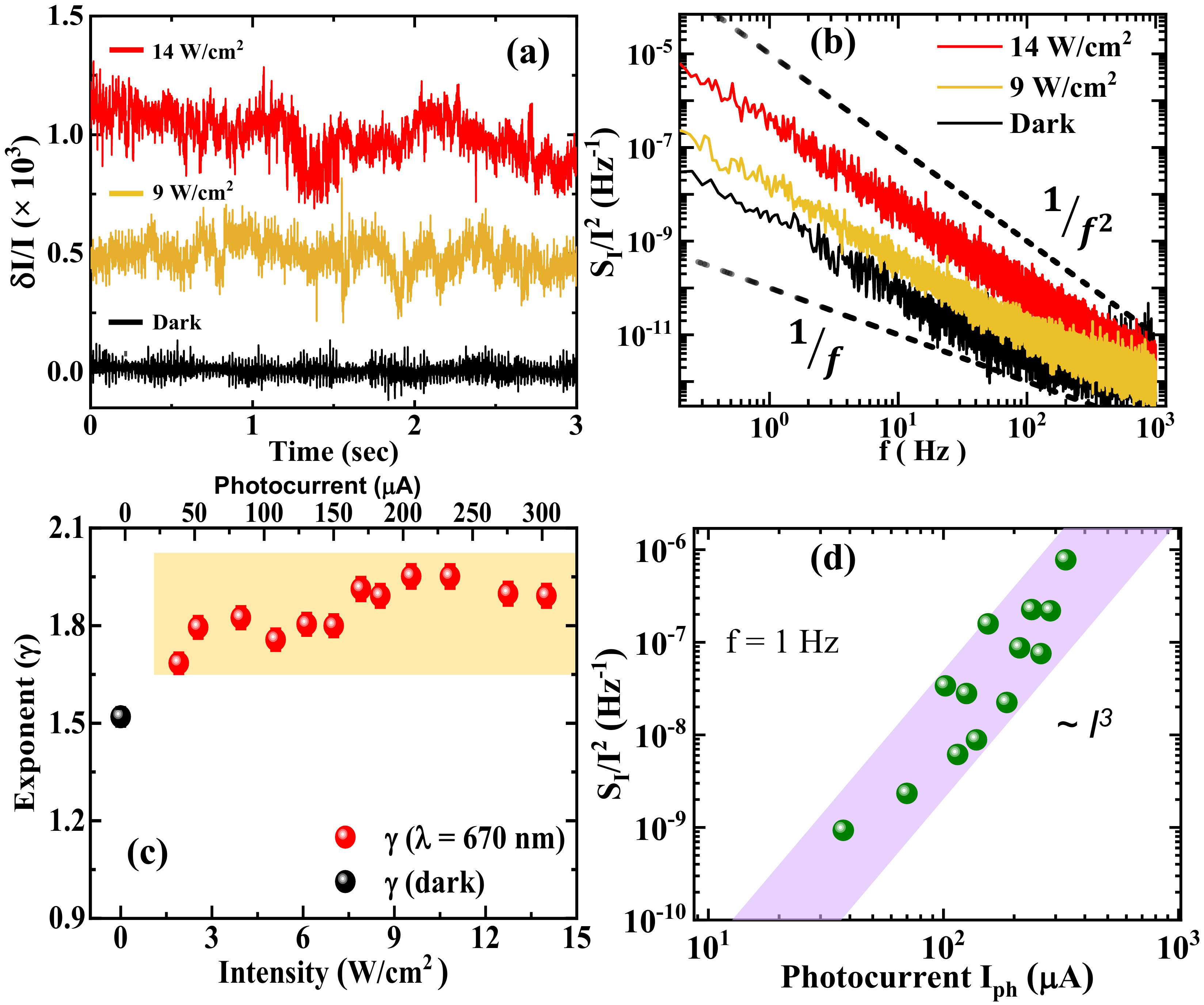}
\caption{\textbf{ 1/f-noise characteristics of the photodetector.} (a) Representative time series of current fluctuations with a vertical shift for clarity. (b) Room temperature normalized Power PSD for laser power. The black dotted lines represent boundaries of $1/f^1$ and $1/f^2$. (c) Variation of noise exponent ($\gamma$) with laser power. (d) Noise-floor at 1 Hz with photocurrent.}
\label{Fig:PhNoise}
\end{figure}

To understand the transport mechanism and impact of laser fluence on the noise level, LFCNS under reverse bias is carried out with laser power 0.1-14 W/cm$^2$. Representative time series of current fluctuations are presented in Fig [\ref{Fig:PhNoise}(a)]. 
%Fig6_Noise2

The PSD of the photo-current fluctuations exhibits typical $1/f^{\gamma}$ behavior [see Fig. \ref{Fig:PhNoise}(b)]. The noise level increases gradually with the increase in intensity which is consistent with the fact that S$_I \propto \langle I \rangle ^2$. Moreover, the exponent ($\gamma$) increases from 1.51 (dark condition) to 2 (see Fig. [\ref{Fig:PhNoise}(c)] ). Such a high value of $\gamma$ indicates significant contributions of carriers having large relaxation time. 1/f noise component of PSD at Hz is shown in Fig. [\ref{Fig:PhNoise}(d)] 

The possible transport mechanisms include processes like absorption, de-absorption of ions, or diffusion. However, such ionic mechanisms cannot give long-time constants. Therefore possible origin of such electronic phenomena can be the trapping and de-trapping of charges across the junction. Trap states that are located close to the barrier potential barrier undergo a continuous birth and death process under photo illumination. Local fluctuation in the trap states during filling and emptying can produce a significant contribution to the excess 1/f noise. 
\noindent
\underline{{\em 1/f noise and photocurrent :}} To sum up, the noise floor of the current fluctuations holds power-law dependence on $I_{ph}$. Such power-law dependence is a characteristic feature of the photodetectors earlier observed in 1980 for HgTe\cite{Tobin1980_HgTe}, later in 2021 for P-InAsSbP/n-InAs\cite{dyakonova2021low}. In the present investigation, we observed that 1/f component of exhibits S$_I\propto I_{ph}^3$.

\section{Conclusions}\label{sec3}
We have fabricated  \ch{Bi2Se3}/Si heterojunction (i.e. p-n diode) based photodetector via a cost-effective chemical route. The junction parameters and optoelectronic transport properties have been benchmarked by temperature-dependent IV and CV measurements. The device exhibits the highest photo responsivity and detectivity of 130 A/W and 3×10$^{10}$ Jones, respectively and shows an ideality factor, $\eta \sim 1.3$ at room temperature. 

The forward noise characteristics exhibit typical 1/f type features with excess noise near the temperatures where a broad hump is observed in the temperature-dependent resistance curve. The uni-slope behaviour in 1/f noise over four decades of frequency suggests a homogeneous distribution of barrier height.

The photodiode exhibits longer time scale of the OFF-state compared to the ON-state in photo-switching possibly due to the contribution from trap-assisted charge transports. Such a trap-de-trapping charge transport mechanism is also corroborated by the noise spectra of photo-generated current fluctuations. The quantification of 1/f noise in current fluctuations establishes a framework for characterizing the hybrid systems to determine the technological applications. This study sheds light on the impact of barrier inhomogeneity and trapped states on device performances. 
\clearpage
\section{Experimental Section}

\subsection{Preparation of \ch{Bi2Se3} nano-flakes}
For the preparation of \ch{Bi2Se3} nano-flakes, a chemical bath technique is imposed. In this chemical route \ch{Bi(NO3)3}.5\ch{H2O} (99.99 $\%$) and \ch{Na2SeSO3} (99.99 $\%$) are used as sources of Bi and Se respectively Where 0.1 M \ch{Bi(NO3)3}.5\ch{H2O} and 80 mL of 0.1 M nitrilotriacetic acid(NTA) are mixed to form bismuth Chelate in a chemical bath. Following chelate formation, a reducing solution of 4 mL of 0.5 M ascorbic acid (AA) is poured into the chelate with continuous stirring. To maintain the pH of the above solution close to 9.0 Ammonia solution is used dropwise and with time the solution turns transparent. Freshly prepared 6 mL 0.1 M \ch{Na2SeSO3} solution is poured into the above mixture.

\subsection{Characterization}

To understand the crystallinity and phase purity, the X-ray diffraction (XRD) pattern of \ch{Bi2Se3} film has been carried out using (Bruker D8 Advanced) with Cu K$_\alpha$ radiation ($\lambda$=1.54056 \AA) with scan range($2\theta$) 15$^{\circ}$- 60$^{\circ}$. Surface morphology and junction cross-section have been studied using a Field Emission Scanning Electron Microscope (FESEM; JEOL; MODEL JSM1T300HR) which is equipped with an energy-dispersive X-ray spectroscopy (EDX). The crystalline quality of the \ch{Bi2Se3} flakes is observed under an Ultra High-Resolution Field Emission Gun Transmission Electron Microscope (UHR-FEG-TEM)(Fig.1e).

\subsection{Device Fabrication}
Boron-doped p-type Si(111) wafers are cut into 2 cm × 1 cm pieces, cleaned, and partially masked. The neatly cleaned wafer is vertically dipped into the as-prepared solution mixture and is kept in the hot oil bath at 75$^{\circ}$ for 1 hour with constant stirring. Deposited films are rinsed carefully with de-ionized water and ethanol to remove unreacted precursors (if any) and kept for drying naturally. Electrode connections are made using enameled copper wire with a dimension of 50 AWG for transport and noise characteristics.

%\backmatter
%\bmsection*{Author contributions}
%\textcolor{red}{S.K. and B.D. performed conceptualization, data acquisition, validation, formal analysis, investigation, data curation, and wrote the original draft, and revised the final manuscript. S.C. performed validation, wrote the original draft, and revised the final manuscript. A.B. and S.B. performed validation and data curation. K.K.C. and M.M. L. performed the investigation and data curation, formal analysis, investigation, data curation, funding acquisition, resources, and wrote the original draft, and revised the final manuscript.}

%\bmsection*{Acknowledgments}
\section{Acknowledgments}
This work was partly supported by the 'Department of Science and Technology, Government of India (Grant No. SRG/2019/000674 and CRG/2023/001100).  S.K. thanks IACS for the Ph.D. fellowship. K.K.C. acknowledges the University Grants Commission, Govt. of India, under the 'University with potential for excellence II' (UPE II) scheme. AB \& SB thanks CSIR Govt. of India for Research Fellowship with Grant No. 09/080(1109)/2019-EMR-I \& 09/080(1110)/2019-EMR-I, respectively.

%\bmsection*{Conflict of interest}
\section{Conflict of interest}
The authors declare no potential conflict of interest.

%\bibliography{achemso-demo}
%\bibliography{Refs_maintext}

\providecommand{\latin}[1]{#1}
\makeatletter
\providecommand{\doi}
  {\begingroup\let\do\@makeother\dospecials
  \catcode`\{=1 \catcode`\}=2 \doi@aux}
\providecommand{\doi@aux}[1]{\endgroup\texttt{#1}}
\makeatother
\providecommand*\mcitethebibliography{\thebibliography}
\csname @ifundefined\endcsname{endmcitethebibliography}  {\let\endmcitethebibliography\endthebibliography}{}

\end{document}